\documentclass[conference]{IEEEtran}
\IEEEoverridecommandlockouts

\usepackage{cite}
\usepackage{amsmath,amssymb,amsfonts}
\usepackage{algorithmic}
\usepackage{graphicx}
\usepackage{textcomp}
\usepackage{xcolor}
\usepackage{array}
\usepackage{booktabs}
\usepackage{makecell}

\newcolumntype{P}[1]{>{\centering\arraybackslash}p{#1}}

\def\BibTeX{{\rm B\kern-.05em{\sc i\kern-.025em b}\kern-.08em
    T\kern-.1667em\lower.7ex\hbox{E}\kern-.125emX}}

\begin{document}


\title{Phase-Aware CNN for Real-Time 5G/6G Channel Estimation with Hardware-in-the-loop Validation}

\author{
\IEEEauthorblockN{
J.~Zolfaghari-Bengar,
R.~Rony,
E.~Gomez-de-Lope,
A.~Villena-Rodriguez,
A.~Mahadevan,
N.~Kourtellis
}
\IEEEauthorblockA{
Keysight AI Labs\\
\{javad.z, rakibul.rony, elisa.gomez-de-lope, alejandro.villena-rodriguez, abhinav.mahadevan, nicolas.kourtellis\}@keysight.com}}


\maketitle

\begin{abstract}
In 5G/6G wireless systems, accurate and timely channel estimation is critical to ensure reliable communication under complex, fast-changing radio conditions. This work focuses on pilot-based channel estimation using deep learning to reconstruct both magnitude and phase across the full subcarrier grid, with particular emphasis on evaluation using emulated data collected from an end-to-end O-RAN testbed. The testbed includes hardware in the loop and controlled channel emulation to better reflect deployment conditions beyond pure software simulation. It addresses major limitations in classical estimators such as LS and MMSE, as well as deep learning-based approaches that struggle with phase prediction due to discontinuities at $\pm \pi$, poor generalization to different UE and antenna configurations, and computational inefficiency for real-time deployment. The proposed system combines a phase-aware input encoding using sine and cosine representations with a lightweight Convolutional Neural Network (CNN) architecture. This design achieves high accuracy, stable phase reconstruction, strong generalization across testbed-derived datasets, and real-time inference suitable for edge devices.
\end{abstract}



\section{Introduction}

Accurate and reliable channel estimation is a fundamental requirement in modern 5G and emerging 6G communication systems. The increasing density of deployments, higher carrier frequencies, and diverse mobility profiles place stringent demands on the ability of user equipment and base stations to track rapidly time-varying wireless channels \cite{b1}. Pilot-based channel estimation plays a central role in enabling robust demodulation, interference management, and beamforming in these systems. However, classical estimation methods face major challenges in realistic environments where fading, noise, and hardware impairments distort pilot signals in complex ways.

The Least Squares (LS) estimator, although computationally efficient, performs poorly in low-SNR regimes and cannot exploit channel statistics. Minimum Mean Square Error (MMSE) estimation improves accuracy but relies heavily on covariance matrices that are often unavailable, outdated, or unreliable in practice \cite{b2}. Moreover, MMSE involves high computational complexity, particularly in massive Multiple Input Multiple Output (MIMO) settings, making real-time operation difficult. These limitations have motivated significant interest in data-driven channel estimators based on deep learning (DL), which have demonstrated promising performance in various Orthogonal Frequency Division Multiplexing (OFDM) and MIMO scenarios \cite{b3, b4, b5, b6}. Despite these advances, DL-based approaches still struggle with critical issues, including sensitivity to phase discontinuities at $\pm\pi$, limited generalization across diverse UE and antenna configurations \cite{b7, b8}, and inference complexity that restricts deployment on edge devices.

A major limitation of existing research is the extensive reliance on synthetic datasets generated through statistical or geometry-based channel models. While such models provide valuable baselines, they fail to capture important real-world phenomena such as hardware nonlinearities, calibration offsets, quantization effects, mutual coupling, and implementation-specific RF impairments. These factors significantly influence the structure of received pilots and channel responses. To address this gap, this work leverages hardware-enabled measurements collected from a fully integrated testbed featuring commercial O-RAN components, multi-antenna RF generation, and programmable channel emulation. Such testbed-derived data reflect the true operational behavior of practical systems and expose learning models to imperfections that are difficult to model or simulate accurately.

\begin{figure*}[htbp]
\centering
\includegraphics[width=\textwidth]{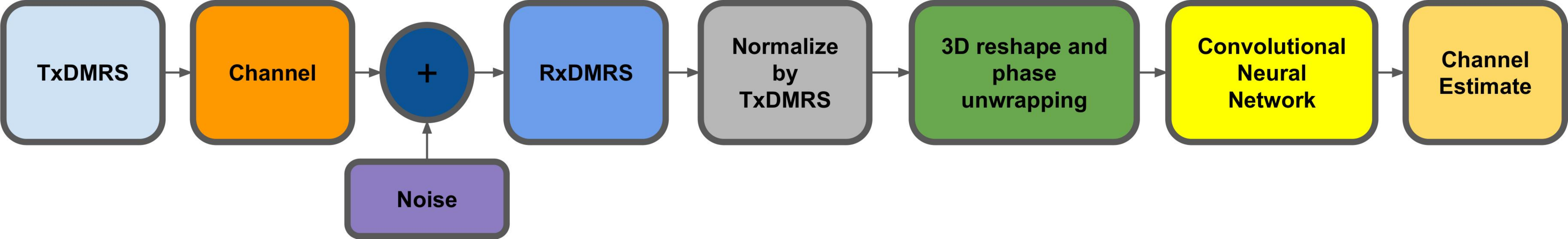}
\caption{Pipeline depicting the process of channel estimation using CNN, starting from the transmission of TxDMRS, through the channel and noise addition, followed by normalization, reshaping, phase unwrapping, and ultimately channel estimation.}
\label{fig1}
\end{figure*}

Motivated by these challenges, this work introduces a novel phase-aware CNN architecture designed to overcome the limitations of both classical and DL-based estimators. By representing phase using continuous sine–cosine features and employing a lightweight convolutional structure with a robust normalization strategy, the proposed system delivers stable magnitude and phase reconstruction across diverse 5G channel environments. Evaluations based on testbed-derived datasets demonstrate that the model achieves strong cross-scenario generalization, improved robustness to hardware distortions, and inference latency suitable for real-time edge deployment \cite{b9, b10, b11}. An overview of the complete channel estimation pipeline, from Transmitted DeModulation Reference Signal (TxDMRS) transmission through normalization and CNN-based reconstruction, is illustrated in Fig.~\ref{fig1}.

\section{Related Works}

\subsection{Classical Channel Estimation}

The LS estimator is commonly used due to its simplicity, but it is highly sensitive to noise and does not incorporate channel statistics \cite{b1}. This leads to poor performance in low-SNR or interference-heavy environments, particularly in high-mobility or millimeter-wave scenarios \cite{b12}. Conversely, MMSE estimation leverages channel and noise covariance knowledge to improve estimation accuracy, but deriving accurate covariance models in real-world deployments is difficult due to hardware impairments and changing propagation conditions. Furthermore, MMSE requires high-dimensional matrix inversion, making it computationally prohibitive for large MIMO configurations \cite{b2}. Extensions such as Bayesian MMSE \cite{b13} and sparsity-aware channel estimation \cite{b14} offer improvements but still rely on assumptions that may not hold in practical systems. Compressive-sensing-based approaches \cite{b15} exploit sparsity in delay–Doppler or angular domains, but performance deteriorates when channels deviate from assumed structures.

\subsection{Deep Learning-Based Channel Estimation}

Deep learning has gained attention as a flexible and data-driven alternative to classical estimators. CNNs, Recurrent Neural Networks (RNNs), transformer-based networks, and hybrid architectures have been proposed for OFDM and MIMO channel estimation \cite{b3,b4,b5,b6,b7,b9,b10}. Despite showing promising gains, existing approaches face several challenges. They often struggle with phase discontinuities at $\pm\pi$, sensitivity to domain shift, and limited generalization across channel models, antenna configurations, or Signal to Noise Ratio (SNR) ranges \cite{b8}. Ye et al. \cite{b5} proposed CNN-based OFDM estimators but did not address phase wrapping issues. Deep unfolding techniques \cite{b16} incorporate model-based structures into neural networks but remain computationally demanding. RNN-based models \cite{b9,b17} capture temporal correlation but are primarily suited for prediction rather than per-subcarrier estimation. Transformer models \cite{b18} better capture long-range dependencies but introduce substantial computational cost.

Beyond OFDM estimation, DL approaches have explored channel modeling and prediction \cite{b7,b19}, uncertainty-aware channel estimation \cite{b11}, and end-to-end neural receivers \cite{b20,b21}. Although these works advance DL for physical-layer processing, most rely heavily on synthetic datasets generated using simplified The 3rd Generation Partnership Project (3GPP) or geometric channel models. As shown in recent studies, DL models trained exclusively on simulated data often fail to generalize in real deployment scenarios due to hardware-induced impairments, nonlinearities, and calibration mismatches not captured in simulations.

To further enhance generalization, recent works have explored meta-learning \cite{b22}, reinforcement learning for channel tracking \cite{b23}, and federated learning for distributed Channel State Information (CSI) prediction \cite{b24}. However, these approaches introduce new challenges such as high computational cost, communication overhead, and instability under non-independent, identically distributed (IID) data distributions. Overall, while the DL literature on channel estimation is rich and rapidly growing, the heavy dependence on simulated datasets and the lack of validation on hardware-enabled testbeds remain major limitations for practical deployment in 5G/6G systems. A concise comparison between conventional estimators and the proposed approach is provided in Table~\ref{tab:comparison}, highlighting key differences in robustness, complexity, and deployability.

\begin{table}[t]
\caption{Summary comparison between our proposed method and conventional approaches}
\label{tab:comparison}
\centering
\small
\setlength{\tabcolsep}{5pt}
\renewcommand{\arraystretch}{1.2}
\begin{tabular}{p{2.2cm} P{1.4cm} P{1.7cm} P{2.0cm}}
\toprule
\textbf{Criterion} \rule{0pt}{3.0ex} &
\textbf{LS Estimation} &
\textbf{MMSE Estimation} &
\textbf{Proposed DL-Based Method} \\
\midrule

\textbf{Accuracy} &
$\times$ &
requires priors &
$\checkmark$ \\

\textbf{Robust to Noise} &
$\times$ &
$\checkmark$ (if covariance accurate) &
$\checkmark$ \\

\textbf{No Need for Channel Priors} &
$\checkmark$ &
$\times$ &
$\checkmark$ \\

\textbf{Inference Time} &
$\checkmark$ &
$\times$ &
$\checkmark$ (30\,ms/batch unoptimized) \\

\textbf{Edge Deployable} &
$\checkmark$ &
$\times$ &
$\checkmark$ \\

\textbf{Generalization} &
$\times$ &
$\times$ &
$\checkmark$ \\

\textbf{Smooth Phase Recovery} &
$\times$ &
theoretically &
$\checkmark$ \\

\bottomrule
\end{tabular}
\end{table}


\section{Methodology}

\begin{table*}[!t]
\caption{Phase-Aware CNN Architecture Used for Channel Estimation}
\centering
\renewcommand{\arraystretch}{1.25}
\begin{tabular}{p{2.0cm} p{2.0cm} p{2.3cm} p{1.8cm} p{1.8cm} p{6.0cm}}
\toprule
\textbf{Layer} &
\textbf{Type} &
\textbf{Filters / Units} &
\textbf{Kernel Size} &
\textbf{Activation} &
\textbf{Notes} \\
\midrule

\textbf{Input} & --- & --- & --- & --- &
Input shape: [batch\_size, subcarriers, UE $\times$ Rx antennas, 3] \\
\textbf{} & & & & &
Input features: $|Y_{\mathrm{DMRS}}|$, $\sin(\angle Y_{\mathrm{DMRS}})$, $\cos(\angle Y_{\mathrm{DMRS}})$ \\

\hline

\textbf{Layer 1} & Conv2D & 32 & (5,1) & ReLU & Same padding \\
\textbf{}        & BatchNorm & --- & --- & --- & --- \\
\hline

\textbf{Layer 2} & Conv2D & 64 & (5,3) & ReLU & Same padding \\
\textbf{}        & BatchNorm & --- & --- & --- & --- \\
\hline

\textbf{Layer 3} & Conv2D & 64 & (3,3) & ReLU & Same padding \\
\textbf{}        & BatchNorm & --- & --- & --- & --- \\
\hline

\textbf{Layer 4} & Conv2D & 32 & (3,1) & ReLU & Same padding \\
\textbf{}        & BatchNorm & --- & --- & --- & --- \\
\hline

\textbf{Layer 5} & Conv2D & 3 & (1,1) & Linear &
Output shape: [batch\_size, subcarriers, UE $\times$ Rx antennas, 3] \\
\textbf{} & & & & &
Output features: $|H|$, $\sin(\angle H)$, $\cos(\angle H)$ \\
\bottomrule
\end{tabular}
\label{tab:cnn_architecture}
\end{table*}










The proposed channel estimation pipeline leverages the structure of DMRS-based uplink while avoiding the limitations of classical LS and MMSE estimators. Following the structure of an uplink system, the UE transmits known pilot sequences (TxDMRS) in the Radio Frequency (RF) domain where they undergo multipath fading, hardware impairments and additive noise before being received as RxDMRS. Instead of forming intermediate estimates based on analytical approximations or requiring prior channel statistics, the pipeline normalizes RxDMRS to match the power of TxDMRS, compensating for variations in transmit power or UE–BS link budget. This produces a stable and consistent input representation that reflects only propagation- and noise-induced distortions, forming the basis for learning the full channel response.

\subsection{CNN Model}

The normalized DMRS tensors are processed by a lightweight CNN designed to capture both spatial and frequency-domain correlations in the MIMO channel. The layer-by-layer configuration—including kernel sizes, filter counts, activations, and tensor shapes—is summarized in Table~\ref{tab:cnn_architecture}. The architecture maintains a small memory footprint (under 1 MB), enabling real-time deployment on edge hardware.

Sequential convolutional layers with residual connections extract hierarchical patterns across antenna elements and pilot (or subcarrier) dimensions. Small kernels along the antenna dimension learn fine spatial variations, while elongated kernels across the frequency dimension capture cross-subcarrier smoothness. Batch normalization layers stabilize feature distributions, improving convergence and reducing overfitting. After feature extraction, the final convolutional and fully connected layers reconstruct the magnitude and phase of the channel in a noise-suppressed and physically consistent manner.

\subsection{Phase Representation}

To ensure stable learning of phase information, the complex channel is converted into a magnitude–phase parameterization, where the phase is encoded using its sine and cosine components. Raw phase values suffer from discontinuities at $\pm\pi$, leading to sharp jumps that destabilize gradient-based learning. The sin–cos encoding eliminates this discontinuity, provides smooth regression targets, and preserves the circular structure of the phase.

Formally, each complex channel coefficient is written as
\begin{equation}
H = |H|e^{j\phi}, \quad \phi = \angle H.
\end{equation}
Instead of directly regressing the wrapped phase value $\phi$, the input and output channel representation are defined as
\begin{equation}
\mathbf{x} = \left[ |H|, \sin(\phi), \cos(\phi) \right].
\end{equation}
During inference, the phase is reconstructed from the predicted sine and cosine components as
\begin{equation}
\hat{\phi} = \operatorname{atan2}\left(\widehat{\sin(\phi)}, \widehat{\cos(\phi)}\right).
\end{equation}
This continuous circular representation avoids the artificial discontinuity at $\pm\pi$ and makes the regression objective smoother than direct phase prediction.

Magnitude, $\sin(\theta)$, and $\cos(\theta)$ are organized into structured 3D tensors spanning antenna dimensions, subcarriers, and feature channels. This representation exposes informative spatial–frequency dependencies to the CNN, enabling it to learn the multipath structure and frequency-selective fading patterns inherent to real channels. The overall preprocessing pipeline is illustrated in Fig.~\ref{fig1}.

\subsection{Robustness to Noise}

Noise robustness is critical for reliable operation in low-SNR or interference-heavy conditions typical of 5G and future 6G systems. Classical LS estimates propagate noise directly, yielding an error variance proportional to the noise variance. In contrast, the CNN leverages data-driven structural priors—smoothness across subcarriers, spatial correlation across antennas, and typical multipath signatures—to suppress uncorrelated noise during inference.

To evaluate noise resilience, an SNR sweep is performed in which ground-truth channels and transmitted pilots are held constant while varying levels of additive Gaussian noise are injected into RxDMRS. This isolates the effect of noise and demonstrates the model’s capacity to maintain estimation quality even at low SNR.

It is worth noting that the hardware-based nature of the data used in this paper imposes a non-negligible noise floor. This is mainly caused by the non-idealities of real RF components and commercial O-RAN equipment.

\begin{figure*}[htbp]
\centering
\includegraphics[width=\textwidth]{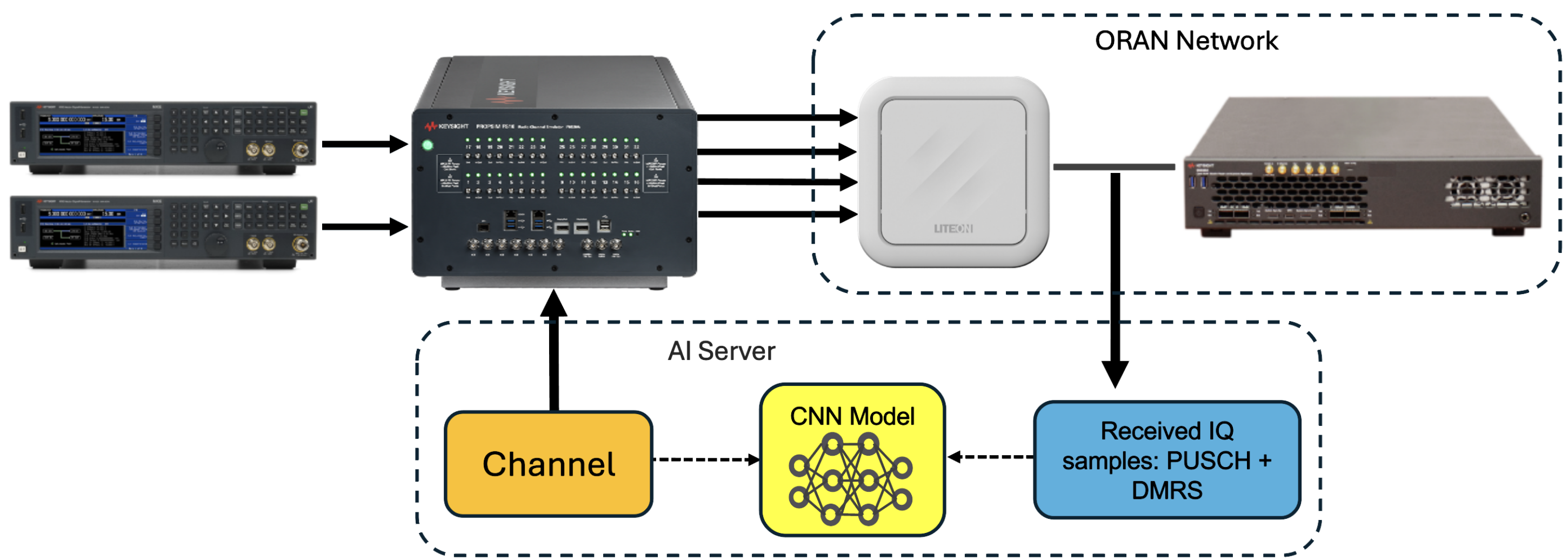}
\caption{Overall testbed setup including RF generators, channel emulator, O-RAN RU, DU emulation, and AI server.}
\label{fig:testbed}
\end{figure*}

\section{Experimental Setup}
\label{sec:experimental_setup}

Tables~\ref{tab:wireless_parameters} and \ref{tab:testbed_parameters} summarize the main wireless-system parameters and hardware implementation details used throughout the experiments.


\subsection{Dataset and Testbed Description}

The dataset used in this study covers a broad range of 5G propagation environments found in the 3GPP standard. To generate diverse enough data, we consider three channel models: Rural Macro (RMa), Urban Macro (UMa), and Urban Micro (UMi). For each of these scenarios two parameters can be modified, namely UE placement (indoor/outdoor deployments) and Line-of-sight conditions (LOS and NLOS). This enables training and evaluation across diverse multipath and scattering characteristics. For each scenario, a total of 12,000 samples were used for training, with 1,000 for validation and 2,000 for testing. Each sample consisted of TxDMRS, RxDMRS, and the corresponding ground-truth channel frequency response. Complex-valued data was converted to magnitude and phase, with the phase further transformed into sine and cosine components. All experiments consider a $2 \times 4$ MIMO uplink configuration, with two independent single-antenna UEs and a single receiver equipped with four antennas. It is worth noting that the DMRS configuration of the UEs are orthogonal thus minimizing interference. Table~\ref{tab:dataset_summary} summarizes the nine representative scenarios included in the dataset, spanning different terrain types, propagation geometries, and visibility conditions.

\vspace{0.5em}
To further validate performance under controlled yet realistic RF impairments, a hardware-in-the-loop testbed was constructed. The complete setup is illustrated in Fig.~\ref{fig:testbed}, showing the integration of RF hardware, channel emulation, and AI processing pipelines. Two Keysight MXG N5182B RF generators emulate independent UEs transmitting uplink DMRS sequences. These signals pass through a PROPSIM F8800B channel emulator configured with imported channel impulse responses matching the above scenarios. The faded waveforms are fed into a commercial Liteon O-RAN Radio Unit performing RF-to-baseband downconversion, after which a Keysight S5040A DU emulator processes the resulting baseband I/Q samples through the 5G uplink chain. All received data, including DMRS and full channel realizations, are captured and stored on an AI compute server responsible for dataset generation, model training, and inference. 

Unlike purely simulated datasets, the hardware-in-the-loop pipeline captures practical RF effects introduced by the RF generators, channel emulator, commercial O-RAN radio unit, DU processing chain, and data-capture path. These effects include calibration mismatches, quantization effects, synchronization imperfections, phase noise, and implementation-specific nonlinearities that are difficult to model accurately using conventional statistical channel models \cite{b25}. Consequently, the evaluation reflects not only propagation-induced channel variations but also implementation-related distortions that affect received DMRS observations in practical deployments. Unless otherwise stated, all quantitative comparisons between LS, LMMSE, and the proposed CNN use the same hardware-derived channel realizations and received DMRS observations under identical preprocessing conditions.

\subsection{Implementation details}
The implementation was carried out using TensorFlow with GPU acceleration to enable efficient training. The model was trained with an MSE loss applied jointly to magnitude, $\sin(\theta)$, and $\cos(\theta)$. The Adam optimizer with a learning rate of 0.001 and a batch size of 32 was employed across 50 epochs. Inference latency was measured on a GPU, resulting in approximately 30~ms per batch in an unoptimized implementation. This latency is reported at batch level and should be interpreted as an implementation measurement rather than a fully optimized PHY-layer deployment latency.

\begin{table}[t]
\caption{Wireless System Parameters}
\label{tab:wireless_parameters}
\centering
\small
\renewcommand{\arraystretch}{1.15}
\begin{tabular}{p{3.1cm}p{4.1cm}}
\toprule
\textbf{Parameter} & \textbf{Value} \\
\midrule
Wireless standard & 5G NR uplink \\
Carrier frequency & 3.5GHz \\
Bandwidth & 100MHz \\
Numerology / Subcarrier spacing & 1 / 30KHz \\
FFT size & 4096 \\
Resource Blocks & 273 \\
Number of evaluated subcarriers & 1638 \\
DMRS Ports &  [0, 2] \\
DMRS Configuration & Type I \\
DMRS Duration & Single symbol \\
DMRS Additional Position & 1 \\
DMRS PUSCH Mapping & Type A \\
DMRS Type A position & 2 \\
Delay Spread & 300 ns \\
MIMO configuration & $1\times4$ uplink \\
Number of UEs & 2 \\
BS receive antennas & 4 \\
Channel scenarios & RMa, UMa, UMi \\
Propagation conditions & LOS/NLOS; Indoor/Outdoor \\
Training / Validation / Test & 12000 / 1000 / 2000 samples per scenario \\
\bottomrule
\end{tabular}
\end{table}

\begin{table}[t]
\caption{Hardware Testbed and AI Implementation}
\label{tab:testbed_parameters}
\centering
\small
\renewcommand{\arraystretch}{1.15}
\begin{tabular}{p{3.1cm}p{4.1cm}}
\toprule
\textbf{Component} & \textbf{Description} \\
\midrule
RF generators & $2\times$ Keysight MXG N5182B \\
Channel emulator & Keysight PROPSIM F8800B \\
Radio Unit & Liteon O-RAN RU \\
DU emulator & Keysight S5040A \\
Training platform & Apple MacBook Pro (M4 Pro, 24 GB unified memory) \\
GPU acceleration & Apple integrated GPU (TensorFlow Metal) \\
Inference latency & $\approx 30$ ms per batch ($B=32$, unoptimized implementation) \\
\bottomrule
\end{tabular}
\end{table}

The LS and LMMSE baselines are implemented using the same received DMRS observations as the CNN. The LS estimate is computed by dividing the received DMRS symbols by the corresponding transmitted DMRS sequence for each subcarrier, receive antenna, and UE. The LMMSE baseline is implemented in the frequency domain. First, LS channel vectors are formed across the subcarrier axis. Then, an empirical frequency covariance matrix $R_{hh} \in \mathbb{C}^{N_{sc}\times N_{sc}}$ is estimated by pooling ground-truth channel realizations across samples, receive antennas, and UEs. The resulting estimator is
\begin{equation}
\hat{\mathbf{h}}_{\mathrm{LMMSE}} = R_{hh}\left(R_{hh}+\sigma_e^2 I\right)^{-1}\hat{\mathbf{h}}_{\mathrm{LS}},
\end{equation}
where $\sigma_e^2$ denotes the effective LS-domain noise variance after RMS normalization and pilot division. Since $R_{hh}$ is obtained from available ground-truth channel realizations, this baseline is an oracle frequency-domain LMMSE reference that provides a strong second-order statistical benchmark rather than a fully deployable receiver implementation.

\begin{table}[htbp]
\caption{Dataset Summary}
\label{tab:dataset_summary}
\centering
\renewcommand{\arraystretch}{1.25}
\begin{tabular}{p{2.4cm} P{1.6cm} P{1.8cm} P{1.5cm}}
\toprule
\textbf{Scenario} & 
\textbf{Model} & 
\textbf{Indoor/Outdoor} & 
\textbf{LOS/NLOS} \\
\midrule
RMa\_indoor  & RMa & Indoor  & --   \\
RMa\_LOS     & RMa & Outdoor & LOS  \\
RMa\_NLOS    & RMa & Outdoor & NLOS \\
UMa\_indoor  & UMa & Indoor  & --   \\
UMa\_LOS     & UMa & Outdoor & LOS  \\
UMa\_NLOS    & UMa & Outdoor & NLOS \\
UMi\_indoor  & UMi & Indoor  & --   \\
UMi\_LOS     & UMi & Outdoor & LOS  \\
UMi\_NLOS    & UMi & Outdoor & NLOS \\
\bottomrule
\end{tabular}
\end{table}

\section{Results}

This section evaluates the proposed phase-aware CNN estimator through a comprehensive set of experiments covering quantitative metrics, qualitative comparisons, cross-scenario generalization, and robustness to noise.

\begin{figure*}[!t]
\centering
\includegraphics[width=\textwidth]{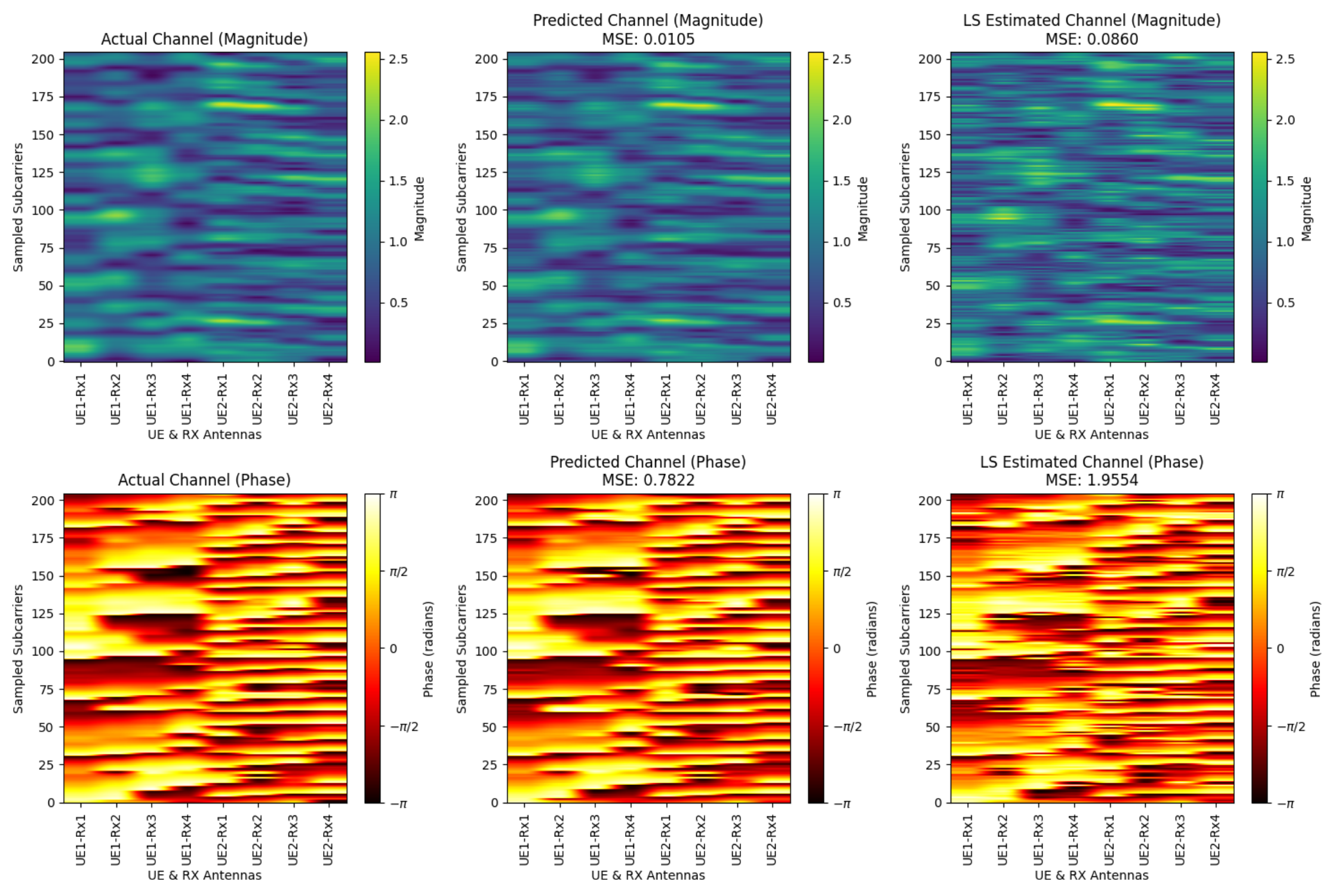}
\caption{Benchmark comparison between the proposed AI-based phase-aware CNN system and the classical Least Squares estimator. The CNN model demonstrates significantly lower errors in both magnitude and phase, producing smoother, more realistic channel reconstructions. The y-axis represents subcarriers sampled at 1 in 8.}
\label{fig2}
\end{figure*}

\begin{figure}[htbp]
    \centering
    \includegraphics[
        width=0.5\textwidth
    ]{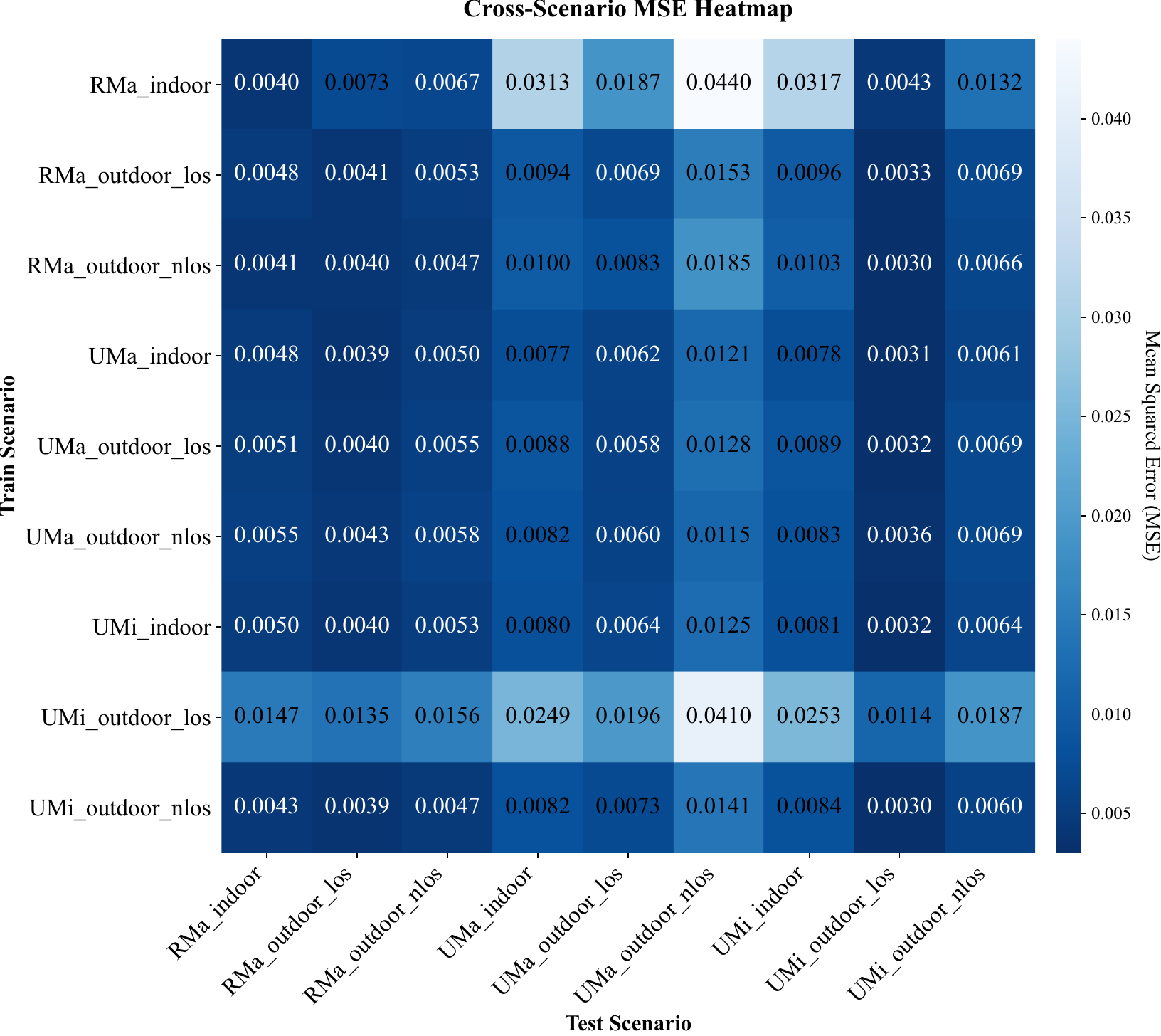}
    \caption{Heatmap of the MSE for cross-scenario model evaluation across Rural Macro (RMa), Urban Macro (UMa), and Urban Micro (UMi) 5G scenarios. Lower values indicate better performance. Diagonal values represent training and testing within the same scenario, whereas off-diagonal values reflect the model's generalization capability across different deployment environments.}
    \label{fig3}
\end{figure}

\begin{figure}[t]
    \centering
    \includegraphics[width=0.9\columnwidth]{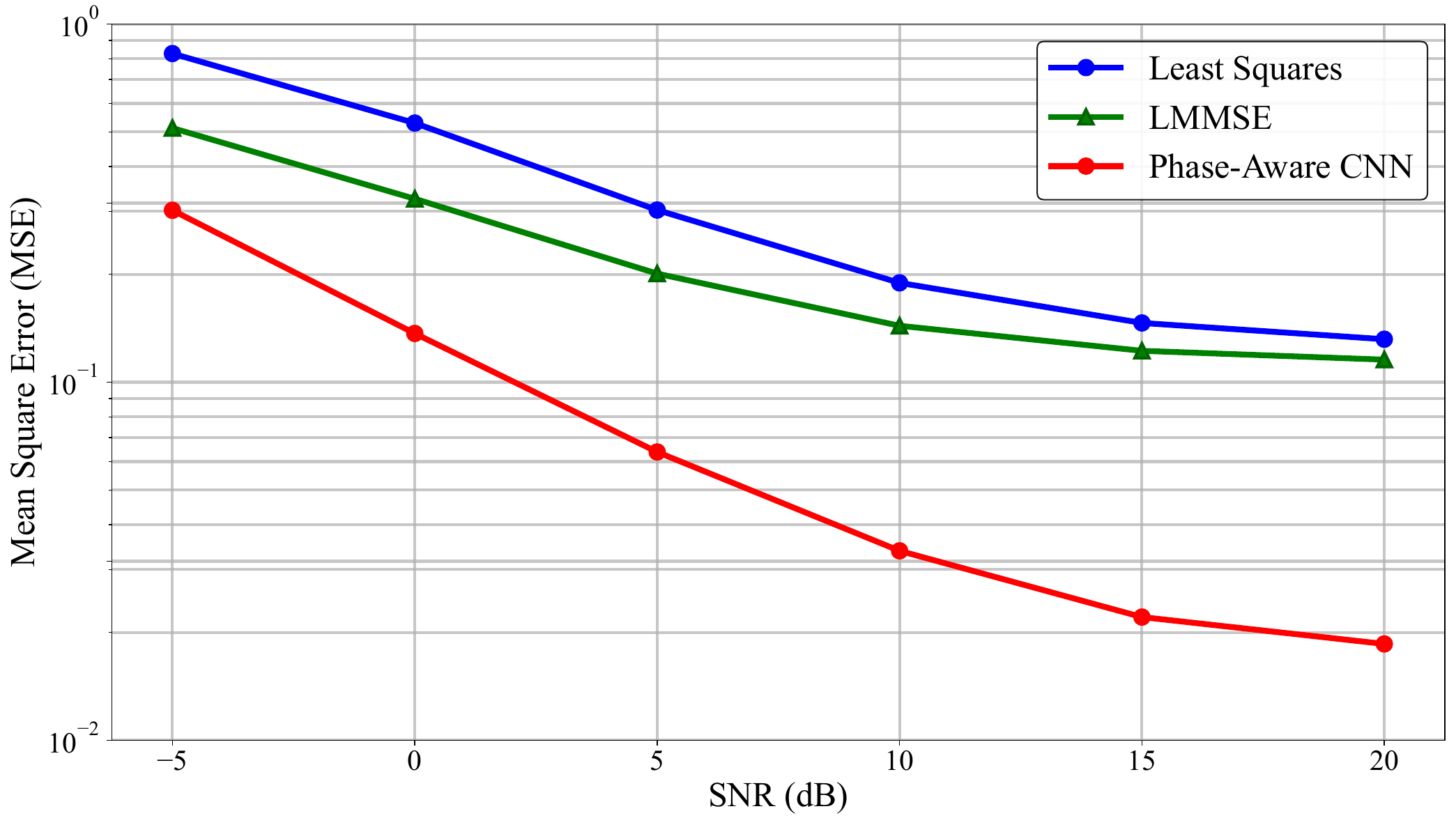}
    \caption{Channel estimation error versus SNR for LS, LMMSE, and Phase-Aware CNN.}
    \label{fig:snr_plot}
\end{figure}

\subsection{Quantitative Results}

The quantitative performance of the proposed CNN estimator is first compared against the classical LS baseline. The resulting magnitude and phase MSE values are summarized in Table~\ref{tab:ls_cnn_results}.

\begin{table}[!t]
\caption{Experimental results when comparing the proposed architecture and Least Squares}
\label{tab:ls_cnn_results}
\centering
\renewcommand{\arraystretch}{1.25}
\begin{tabular}{p{2.0cm} P{2.0cm} P{2.0cm}}
\toprule
\textbf{Estimator} & 
\textbf{Magnitude MSE} & 
\textbf{Phase MSE} \\
\midrule

\textbf{Phase-Aware CNN} & 0.0105 & 0.7822 \\
\hline

\textbf{LS} & 0.0860 & 1.9554 \\
\bottomrule
\end{tabular}
\end{table}

The proposed CNN demonstrates significant quantitative improvements over classical LS estimation. The model’s magnitude MSE of 0.0105 represents nearly an order of magnitude improvement over LS, which yields an error of 0.0860. In the phase domain, the CNN achieves a phase MSE of 0.7822, substantially outperforming the LS error of 1.9554. These results highlight the model's ability to suppress noise and reconstruct smoother magnitude and phase profiles across the entire frequency grid. 

\subsection{Qualitative Results}
A representative visual comparison between LS and the proposed CNN is shown in Fig.~\ref{fig2}. Visual comparisons further demonstrate the superiority of the proposed model. LS-estimated channels exhibit noisy, erratic behavior with sharp fluctuations across subcarriers, while the CNN produces highly coherent magnitude and phase structures. The CNN's outputs correspond closely to the smooth multipath profiles expected in realistic wireless channels. These qualitative improvements are especially important for downstream tasks such as beamforming and scheduling that depend on stable and physically meaningful channel estimates.

\subsection{Generalization Results}

As illustrated in Fig.~\ref{fig3}, models trained on UMa channels exhibit stronger generalization to both RMa and UMi scenarios. This behavior can be attributed to the richer multipath structure of urban macro environments, which provides more diverse training samples. In contrast, models trained on UMi LOS conditions show reduced generalization to other environments due to their limited propagation diversity. These results highlight the importance of training on multipath-rich scenarios to improve robustness across real-world deployment conditions.

\subsection{Robustness Against Noise}

The robustness of the proposed CNN estimator is evaluated across a wide range of SNR conditions. Fig.~\ref{fig:snr_plot} compares its MSE performance against LS and MMSE estimators. The reported MMSE curve corresponds to the oracle frequency-domain LMMSE baseline described in Section~\ref{sec:experimental_setup}, using the same hardware-derived channel realizations and received DMRS observations as the CNN. While all methods exhibit decreasing error with increasing SNR, the proposed CNN consistently achieves the lowest MSE across the entire SNR range.

The performance gain is particularly pronounced at low and moderate SNRs, where the CNN significantly outperforms both LS and MMSE. These results confirm that the proposed model learns noise-resilient representations that remain effective under both noise-limited and high-SNR conditions.


\section{Conclusion}

This work presents a lightweight, phase-aware CNN for real-time channel estimation in 5G and future 6G systems. By combining magnitude features with sine–cosine phase encoding, the proposed model avoids phase discontinuities and enables stable, physically consistent learning. The resulting architecture significantly outperforms classical LS and MMSE estimators in both magnitude and phase accuracy, while remaining suitable for edge deployment.

The model is validated using hardware-in-the-loop measurements generated from an end-to-end O-RAN testbed together with controlled channel emulation covering diverse 3GPP propagation scenarios. Experiments across diverse 3GPP channel environments further confirm strong generalization, particularly when trained on multipath-rich urban macro scenarios. In addition, noise robustness studies across a wide SNR range demonstrate that the network learns noise-resilient spatial–frequency representations, consistently maintaining low estimation error under degraded conditions. These results position the proposed approach as a practical and deployment-ready solution for AI-native wireless communication systems.


\end{document}